\documentclass[prl,twocolumn,superscriptaddress,nofootinbib,noshowpacs,preprintnumbers,longbibliography,floatfix]{revtex4-2}
\usepackage[normalem]{ulem}
\usepackage{color} 
\usepackage{bm}
\usepackage{dcolumn}
\usepackage{times}
\usepackage{amssymb} 
\usepackage{amsmath} 

\usepackage[normalem]{ulem}

\usepackage{ragged2e}
\usepackage{graphicx}
\usepackage{epstopdf}
\usepackage[english]{babel}
\usepackage{mathrsfs}
\usepackage{textcomp}
\usepackage{amsthm}
\usepackage{amsfonts}
\usepackage[nice]{nicefrac}
\usepackage{soul}
\usepackage{braket} 
\usepackage[colorlinks=true,citecolor=blue,linkcolor=blue,urlcolor=blue]{hyperref}
\usepackage{cleveref}
\usepackage{lipsum}
\newcommand{\be}{\begin{equation}}
\newcommand{\ee}{\end{equation}}
\newcommand{\bea}{\begin{eqnarray}}
\newcommand{\eea}{\end{eqnarray}}
\newcommand{\ba}{\begin{array}}
\newcommand{\ea}{\end{array}}

\begin{document}

\title{Geometrical frustration, power law tunneling and nonlocal gauge fields from scattered light}

\author{Pavel P. Popov}
\email{pavel.popov@icfo.eu}
\affiliation{ICFO - Institut de Ci\`encies Fot\`oniques, The Barcelona Institute of Science and Technology, 08860 Castelldefels (Barcelona), Spain}

\author{Joana Fraxanet}
\affiliation{ICFO - Institut de Ci\`encies Fot\`oniques, The Barcelona Institute of Science and Technology, 08860 Castelldefels (Barcelona), Spain}

\author{Luca Barbiero}
\affiliation{Institute for Condensed Matter Physics and Complex Systems,
DISAT, Politecnico di Torino, I-10129 Torino, Italy}

\author{Maciej Lewenstein}
\affiliation{ICFO - Institut de Ci\`encies Fot\`oniques, The Barcelona Institute of Science and Technology, 08860 Castelldefels (Barcelona), Spain}
\affiliation{ICREA, Pg. Lluis Companys 23, ES-08010 Barcelona, Spain}

\date{\today}

\begin{abstract}

Designing the amplitude and range of couplings in quantum systems is a fundamental tool for exploring a large variety of quantum mechanical effects. Here, we consider off-resonant photon scattering processes on a geometrically shaped molecular cloud. Our analysis shows that such a setup is properly modeled by a Bose-Hubbard Hamiltonian where the range, amplitude and sign of the tunneling processes of the scattered photonic modes can be accurately tuned. Specifically, by varying the molecular distribution, we demonstrate that different configurations characterized by geometrical frustration, long-range power law hopping processes, and nonlocal gauge fields can be achieved. 
Our results thus represent a powerful and alternative approach to perform an accurate Hamiltonian engineering of quantum systems with non trivial coupling structures.

\end{abstract}
\maketitle

\paragraph*{Introduction---}
Nonlocal couplings~\cite{Defenu2023} are key ingredients to explore quantum mechanical effects which can deeply influence the entanglement~\cite{Eisert2013,Gong2017}, symmetry ~\cite{Bruno2001,Maghrebi2017}, and out-of-equilibrium~\cite{Hauke2013,Richerme2014} properties of complex quantum systems. In addition, when specifically shaped nonlocal couplings can generate the celebrated phenomenon of geometrical frustration ~\cite{Lhuillier2001}. As known, the latter can have the fundamental role of allowing for the appearance of interesting states of matter characterized by various kind of spontaneously symmetry breakings \cite{Nersesyan1998,Lacroix2011} or topological order \cite{Fujimoto2009}. 
Interestingly, alternative and fundamental resources to achieve frustration or, generally, quantum phases with broken symmetries or topological order are classical gauge fields. These latter, in addition to representing one of the most fundamental concepts in quantum mechanics \cite{Aharonov1959,Berry1984}, are a timely research subject \cite{Aidelsburger2018} for different communities ranging from ultracold atomic systems \cite{Spielman2009,Lin2009,Aidelsburger2011,Dalibard2011,Struck2012,Hauke2012,Goldman2014,Celi2014} and solid state \cite{VOZMEDIANO2010109} to photonic \cite{Joannopoulos2011,Umucal2011,HAFEZI2014} and mechanical systems \cite{Aspelmeyer2014,Huber2016,Walter_2016}.\\In this paper, we demonstrate that the three fundamental aspects—long-range couplings, geometrical frustration, and classical gauge fields—can be engineered through the controlled scattering of photons within a carefully designed molecular geometry.
\begin{figure}[t]
\includegraphics[width=\columnwidth]{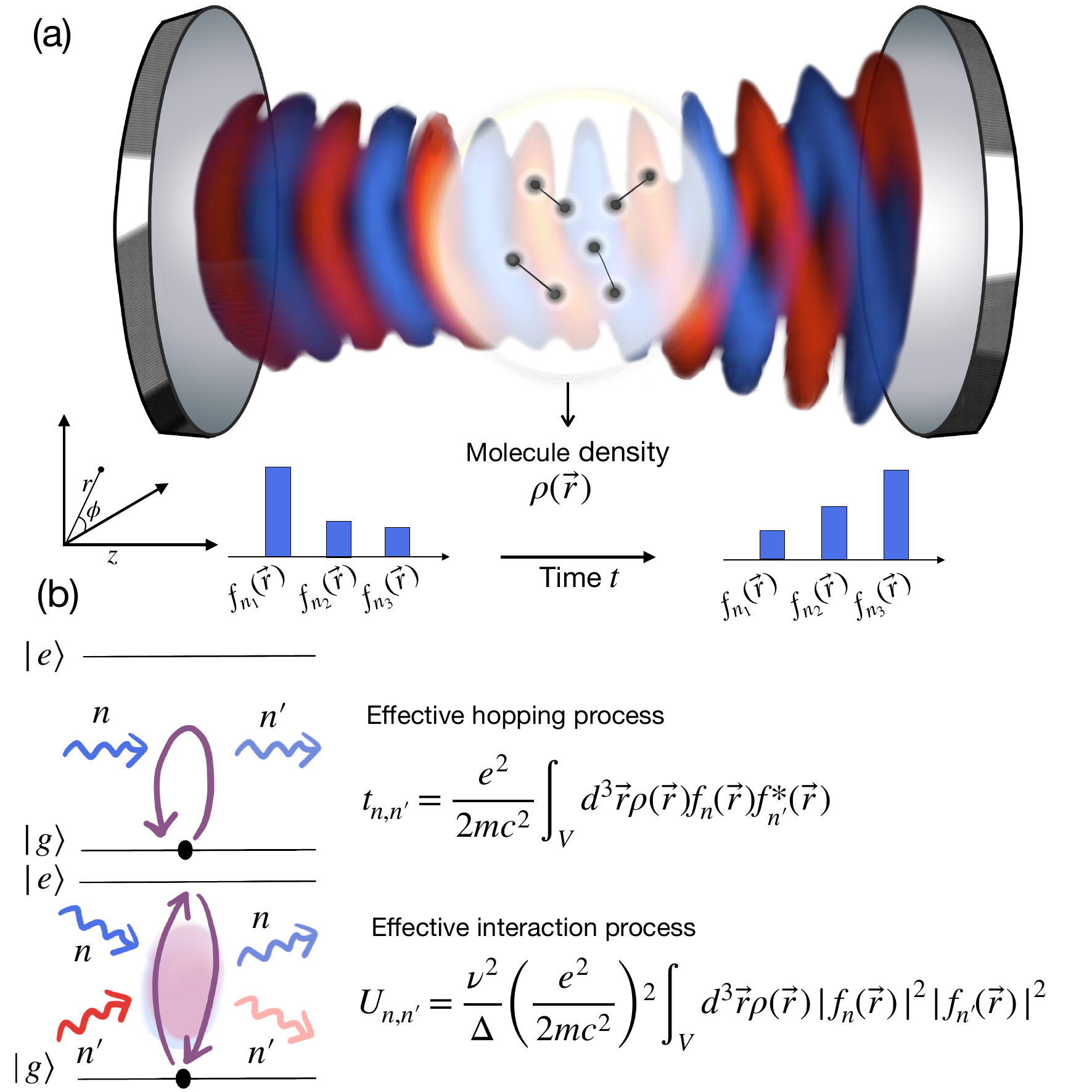}
\caption{\textbf{Physical setup.} (a) A Gaussian beam represented by blue and red patterns illuminates a geometrically shaped molecular cloud in a cavity with molecule density $\rho(\vec{r})$. The beam frequency is tuned so that off-resonant scattering is the dominant light-matter interaction. Off-resonant light scattering onto the molecules induces a dynamic evolution of the beam’s mode occupations $f_{n}(\vec{r})$, e.g., in the Laguerre-Gaussian basis.
(b) To second-order perturbation theory, the effective bosonic Hamiltonian features tunable long-range hoppings $t_{n,n'}$ and density-density interactions $U_{n,n'}$. The geometry of the molecular cloud controls the sign and phase of the hopping and interaction amplitudes.}
\label{fig0} 
\end{figure}
Specifically, we first design a setup where light in form of a Gaussian beam is scattered off resonantly from a molecular cloud. This scheme allows one to derive an effective Bose-Hubbard Hamiltonian \cite{Dutta_2015,chanda2024} where the scattered modes represent the elementary constituents of the considered physical systems. Notably, in such setup the sign, amplitude, and range of the interparticle couplings can be accurately tuned by shaping the geometry of the molecular cloud. Contrary to Ref.~\cite{Asban2019} where the control parameter is the size of the molecular cloud, we utilize different distribution profiles and find that homogeneous lattice geometries can be efficiently realized. Within this approach, we first prove that the scattered modes can be confined to an effective triangular ladder geometry  characterized by geometrical frustration. We then show that a similar manipulation of the molecular distribution generates  highly nonlocal terms in the form of long-range hopping processes of the the scattered modes. Here, we can indeed achieve regimes where the tunneling amplitude of scattered modes follows a power law decay whose decay exponent can be tuned to basically any value. Notably, this feature represents a key aspect for quantum information purposes \cite{Lewis2021,king2025}. Finally, we show that when the profile of the molecular cloud is rotated with respect to the profile of the incident beam, gauge fields affecting the dynamic of the scattered modes can be generated. Notably, we demonstrate that our scheme not only makes it possible for an accurate tuning of the fields value but, thanks to higher nonhomogeneity of the molecular cloud, the gauge fields can also become highly nonlocal, i.e., they can extend over different effective triangular plaquettes thus proving, even more, the relevance of our results.\\
\paragraph{Physical setup and Hamiltonian engineering---}

The setup, as depicted in Fig.~\ref{fig0} (a),  consists of a cavity where the molecules are confined in a cylindrical geometry, with a density distribution $\rho({\vec{r}})$. A single frequency Gaussian beam $\omega_0$ is focused on the molecular cloud, so that off-resonant scattering of the photons on the molecules takes place. The frequency of the photons $\omega_0$ can be tuned far away from typical electronic transitions, so that only off-resonant processes are relevant.
We further assume that the light interacts with the valence electron in each molecule. The minimal Hamiltonian that captures all these processes reads~\cite{Weight2023} 
\begin{align}\label{eq:H_molecular}
{\cal H}  = &\sum_\alpha \frac{1}{2m}\left(\vec{p}_\alpha - \frac{e}{c} \vec{A}(\vec{r}_\alpha)\right)^2 + V(\vec{r}_\alpha, \vec{R}_\alpha)\notag\\ +&  \sum_n \omega_n b_n^\dagger b_n,
\end{align}
where $m, e$ are the electron mass and charge and $c$ is the velocity of light. In Eq.~\eqref{eq:H_molecular}, $V(\vec{r}_\alpha, \vec{R}_\alpha)$ is the potential each electron feels in the molecule, $\vec{r}_{\alpha}$ is the position operator of the electron, $\vec{p}_{\alpha}$ is the corresponding canonical momentum, and $\vec{R}_{\alpha}$ is the position of the molecule. In this expression, the index $\alpha$ is enumerating the different molecules, and the index $n$ accounts for the different photon modes in the expansion of the electromagnetic (EM) part of the Hamiltonian. Since we consider a Gaussian beam, the expansion has to be performed in a set of eigenmodes of the paraxial Helmholtz equation, for more details, see Supplemental Material~\cite{supplemental}. Furthermore, in Eq. \eqref{eq:H_molecular} $\vec{A}(\vec{r}_{\alpha})$ denotes the vector potential of the EM field and $b_n$ is the annihilation operator of the mode with index $n$.  Importantly, here the energy of each mode is given by $\omega_n = \omega_0 + \delta_n$, where $\delta_n$ are small corrections to the frequency of the Gaussian beam.
Neglecting the term $\vec{p}_\alpha\cdot\vec{A}(\vec{r}_\alpha)+\vec{A}(\vec{r}_\alpha)\cdot\vec{p}_\alpha$ in Eq.~\eqref{eq:H_molecular}, as the frequency of the incoming beam is chosen to be off resonant with the electronic transitions in the molecule, we rewrite the Hamiltonian as follows:
\begin{align}\label{eq:H_rewrite}
{\cal H}  = &\underbrace{\sum_\alpha \frac{1}{2m}\vec{p}_\alpha^2 + V(\vec{r}_\alpha, \vec{R}_\alpha)}_{H_0}\notag\\
+& 
\underbrace{\sum_\alpha\frac{e^2}{2mc^2}\vec{A} (\vec{r}_\alpha)^2 + \sum_n \delta_n b_n^\dagger b_n}_{H_1},
\end{align}
where we have dropped constant terms. The wave function of the electronic structure in each molecule can be obtained from the first part of the Hamiltonian, $H_0$. We approximate this structure by a two-level system whose ground state $\ket{g_{\alpha}}$ and excited state $\ket{e_{\alpha}}$ are separated by an energy gap $E = \Delta$ and we assume that all molecules are prepared in their ground state. The second part of the Hamiltonian, $H_1$, describes the evolution of the mode decomposition of the Gaussian beam, when interacting with the scatterers, and can be treated as a perturbation of $H_0$.\\
The first-order correction to the unperturbed part $H_0$ comes from the projection of $H_1$ onto the ground state of $H_0$, which is given by the product,

\begin{align}
    \ket{G} = \prod_\alpha \ket{g_\alpha}\otimes\ket{N},
\end{align}
where we assumed that the total number of photons $N$ does not change (no photon losses)~\cite{RevModPhys.85.553}. Therefore, the first order correction is 
\begin{align}\label{eq:first_order_H}
    \bra{G}H_1\ket{G} = \frac{e^2}{2mc^2}\sum_{\alpha}\bra{g_{\alpha}}\vec{A}(\vec{r}_\alpha)^2\ket{g_{\alpha}}.
\end{align}
The second-order perturbation term is a contribution from a virtual transition to the excited state of the electron in each molecule, which reads 

\begin{align}\label{eq:second_order_H}
&-\frac{1}{\Delta}\sum_{\alpha}|\bra{G}H_1\ket{E_{\alpha}}|^2\notag\\ &= -\frac{1}{\Delta}\bigg(\frac{e^2}{2mc^2}\bigg)^2\sum_{\alpha}|\bra{g_{\alpha}}\vec{A}(\vec{r}_\alpha)^2\ket{e_{\alpha}}|^2,
\end{align}
where we denoted as $\ket{E_\alpha}$ the state in which the $\alpha$th molecule is excited and all the other molecules are not. In order to simplify the calculations, we rewrite the off-diagonal element as proportional to the diagonal one
\begin{align}\label{eq:off_diagonal_element}
    |\bra{g_{\alpha}}\vec{A}(\vec{r}_\alpha)^2\ket{e_{\alpha}}|^2 = \nu^2 \bra{g_{\alpha}}\vec{A}(\vec{r}_\alpha)^2\ket{g_{\alpha}}^2
\end{align}
where $\nu$, depending on the considered molecules~\footnote{Here we are interested in presenting a general scheme applicable to different kinds of molecules so we do not focus on a specific value of $\nu$.}, relates the ground and the excited state wave functions.
%
Because of the considered cylindric symmetry, a suitable choice to expand the vector potential is represented by Laguerre-Gauss (LG) polynomials ~\cite{Forbes2016} acting as the basis for the photon modes.
These latter have two quantum numbers, the azimuthal index $l$ and the radial index $p$, thus $n$ becomes a multi-index: $n \sim (l,p)$. The vector potential expressed in this basis is 
\begin{align}\label{eq:vector_potential}
    \vec{A}(\vec{r},t) = \sum_{l,p}\vec{\epsilon} f_{l,p}(\vec{r},t)e^{i\omega_0(z/c-t)}b^{\dagger}_{l,p} + \text{H.c.},
\end{align}
where $\vec{\epsilon}$ is the polarization vector (e.g. for circular polarization, $\vec{\epsilon} = (\vec{x}-i\vec{y})/\sqrt{2}$), $b_{l,p}$ is the annihilation operator and $f_{l,p}(\vec{r},t)\approx f_{l,p}(\vec{r})$ ($\delta_{n}$ are small) is the spatial dependence of the $(l,p)$th LG mode (see Supplemental Material for details). Substituting Eq.~\eqref{eq:vector_potential} in Eq.~\eqref{eq:first_order_H} and in Eq.~\eqref{eq:second_order_H} yields the effective time-independent Bose-Hubbard Hamiltonian \cite{Dutta_2015,chanda2024} for the transverse LG modes,
\begin{align}\label{eq:H_eff_MB}
    {\cal H}_{\text{eff}} = &\sum_n \mu_n b^{\dagger}_nb_n + \sum_{n,n'}[|t_{n,n'}|e^{i\theta_{n,n'}}b_{n}b^{\dagger}_{n'}+\text{H.c.}]\notag \\-&\sum_{n,n'}U_{n,n'}[3b^{\dagger}_{n}b_{n} + 4b^{\dagger}_nb_nb^{\dagger}_{n'}b_{n'}], 
\end{align}
where the coefficients $\mu,t$ and $U$ correspond to on-site chemical potential, hopping amplitudes and the strength of density-density interactions, respectively. Crucially, all these terms depend on the geometry of the molecular cloud as follows 
\begin{subequations}
\begin{align}
    \mu_n = \frac{e^2}{2mc^2}\int_{V}d^3\vec{r}\rho(\vec{r})|f_{n}(\vec{r})|^2 + \delta_n,
\label{eq:mu_integral}
\end{align}
\begin{align}
    t_{n,n'} = \frac{e^2}{2mc^2}\int_{V}d^3\vec{r}\rho(\vec{r})f_{n}(\vec{r})f^*_{n'}(\vec{r}),
\label{eq:t_integral}
\end{align}
\begin{align}
    U_{n,n'} = \frac{\nu^2}{\Delta} \bigg(\frac{e^2}{2mc^2}\bigg)^2\int_{V}d^3\vec{r}\rho(\vec{r})|f_{n}(\vec{r})|^2|f_{n'}(\vec{r})|^2.
\label{eq:u_integral}
\end{align} 
\label{eq:couplings}
\end{subequations}
In Supplemental Material, we give an estimation of the magnitude of hoppings and interactions under realistic assumptions on laser and molecular properties. Note that we only considered the photon number preserving terms, since the others oscillate with frequencies 2$\omega_0$ and $4\omega_0$ and therefore average out in the rotating frame on the timescale of the coherent dynamics ($\Delta<\hbar\omega_0$). This relaxes the necessity for high-fidelity photonic Fock state preparation, as postprocessing would be enough for extraction of the relevant information about the dynamics.
It is now important to remark the key aspects characterizing Eq.~\eqref{eq:H_eff_MB}.
First, the hopping coefficients $t_{n,n'} = |t_{n,n'}|e^{i\theta_{n,n'}}$ are in general complex numbers. This is, as we show below, crucial for engineering frustration and effective classical gauge fields.
Second, the interaction processes induced by the second order expansion can be either attractive  when perturbation theory is performed around the molecular ground state, as in Eq.~\eqref{eq:H_eff_MB}, or repulsive by considering perturbation theory around the molecular excited state $\prod_{\alpha}\ket{e_{\alpha}}$.
Third, the prefactor $\nu^2/\Delta$ of the interacting terms involves characteristics that depend on the type of molecules used as scatterers. Therefore, potentially different interaction 
regimes can be explored by considering suitable molecules. 

Before proceeding with studying the versatility of our proposal for realizing extended Bose-Hubbard models, we comment on its experimental feasibility. As we show in Supplemental Material, the order of magnitude of the kinetic as well as of the interaction processes can be tuned, under realistic assumptions, to be similar to the laser frequency, thus suppressing losses and decoherence during the time evolution in high-finesse cavity setups~\cite{zhang2025}. Furthermore, we consider molecules as scatterers due to their rich and dense vibrational spectrum. Sizable interactions can be achieved by choosing a low-energy Raman transition ($\Delta < \hbar\omega_0)$. However, alternative realizations could be considered, e.g., with Rydberg or hyperfine transitions in atoms with similar low frequencies.  
\begin{figure}[t]
    \includegraphics[width=\columnwidth]{
    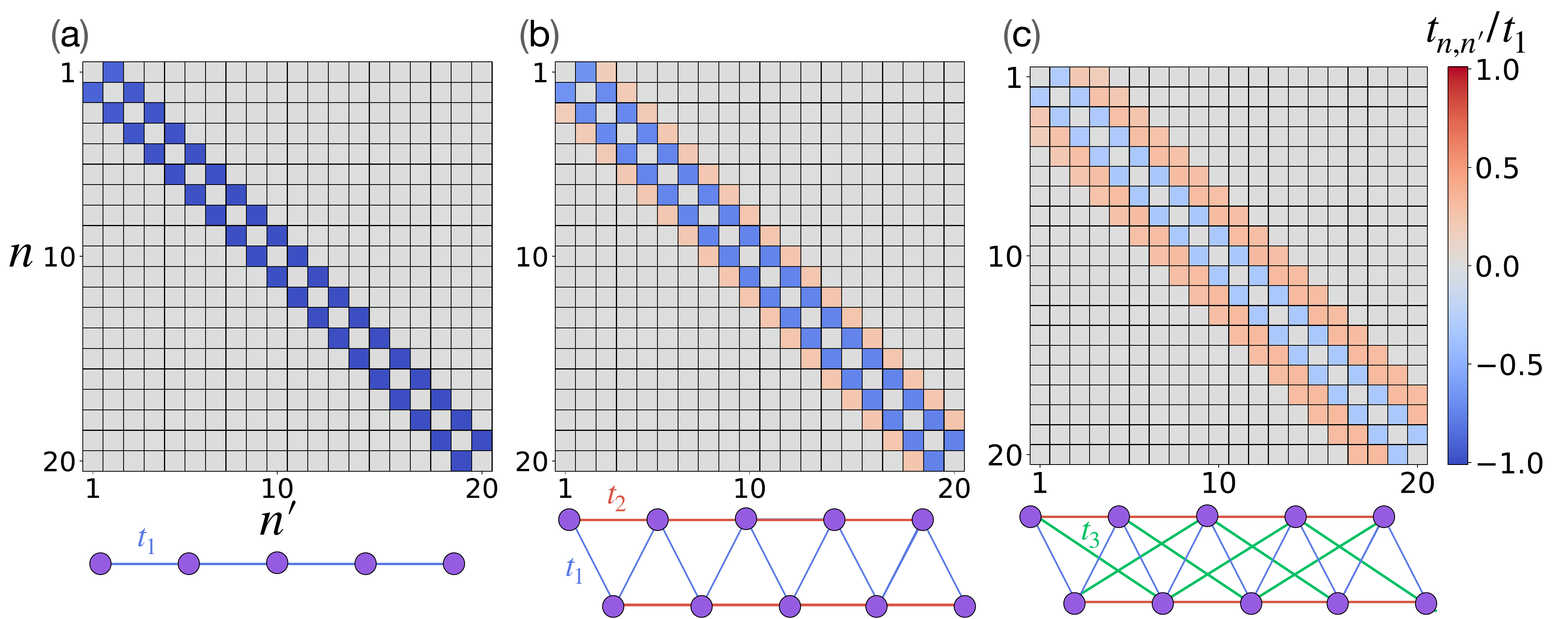}
    \caption{\textbf{Effective geometries.} (a) Purely 1D geometry induced by fixing $c_1=1$, $\varphi_1 = 0.9\pi$, and $c_i=0$ for $\forall i>1$. (b) Effective frustrated triangular geometry obtained by fixing $c_1=3/4$, $c_2=1/4$, $\varphi_1 = 0.9\pi$, $\varphi_2 = 1.1\pi$ and $c_i=0$ for $\forall i>2$. (c) Effective frustrated triangular geometry with hopping amplitudes connecting up to three sites obtained by fixing $c_1=1/3$, $c_2=1/3$, $c_3=1/3$, $\varphi_1 = 0.9\pi $, $\varphi_2 = 1.1\pi$, $\varphi_3 = 1.02\pi$ and  $c_i=0$ for $\forall i>3$.}
    \label{fig1}
\end{figure}
\paragraph{Manipulation of couplings by geometrical shaping---}
In this Section, we demonstrate the high level of versatility encoded in the effective Bose-Hubbard Hamiltonian in Eq. \eqref{eq:H_eff_MB}. Our particular choice for the density of the molecular cloud as a function of the position in cylinder coordinates $\vec{r} = (r,\phi,z)$ reads
\begin{align}\label{eq:density_distribution}
    \rho(r,\phi,z) & = \theta(r)\theta(R-r)\sum^{N_c}_{k = 0}c_k \cos(k\phi+\varphi_k)\nonumber\\
    & \times\theta(z-z_1)\theta(z_2-z).
\end{align}
Here, the amplitudes $c_k$, the phases $\varphi_k$ and the number of cosine terms $N_c$ are free parameters while the spatial extent along the $z-$direction $z_1-z_2$ and the radial extent $R$ are set by the size of the cavity in which the molecules are confined.
Such density profiles can be engineered by employing optical trapping potentials~\cite{bloch2005ultracold,Bloch2008}. Specifically, the special intensity profiles of the LG modes could be of use in shaping $\rho(\vec{r})$. For coarse-grained approximations, one could also use discretely distributed scatterers trapped in optical tweezes.
The simple $\phi$ dependence of the LG modes, $f_{l,p}(\phi)\sim \exp(-il\phi)$, allows us to choose a molecular density as in Eq.~\eqref{eq:density_distribution} that decouples LG modes with different $p$ number. Therefore, if the Gaussian beam is initialized in the $p = 0$ sector, it remains confined in this sector during time evolution. In what follows, we enumerate the modes through their azimuthal index, $n\equiv (l,0)$.\\
Importantly, the coefficients $c_k$ and $\varphi_k$ in the $\phi-$dependence of $\rho(\vec{r})$ serve as control parameters for the hopping amplitudes $t$ and the interaction strengths $U$. In this regard, there are few key observations that need to be made. First, each $c_k$ controls the hopping amplitude between $n$th and $(n+k)$th mode for all $n$ under consideration, thus assuring translational invariance of the effective Hamiltonian. This feature is in crucial difference to previous proposals, where translational invariance is much harder to control~\cite{Asban2019,Katz2022}. Second, the number $N_c$ of nonzero coefficients $c_k$ sets the range of the hoppings, allowing also for up to $N_c$th nearest neighbor, i.e. all to all, hoppings. Third, the complex phase of each hopping amplitude $\theta_{n,n+k}$ is fully controlled by the phase $\varphi_k$. Fourth, the interaction strengths $U$ are controlled also by $c_k$, but are independent of $\varphi_k$. 
To this end, we note that the non-negativity of the density of molecules $\rho(r,\phi,z)\geq 0$ in any point of space imposes constraints on the sum of coefficients $\sum_kc_k$. In the case $\varphi_k = 0 \:\:\forall k$, the constraint is simply $\sum_{k\geq 1}c_k \leq 1$. 
%
In the next paragraphs, we show that Hamiltonians with nontrivial interparticle couplings can be engineered by simply tuning specific coefficients $c_k$.
\paragraph{Effective geometrical frustration---} 
Within the described approach, an extended Bose-Hubbard model featuring nearest neighbor hopping $t_1$ only, see Fig~\ref{fig1}(a), arises by choosing the coefficients in Eq.~\eqref{eq:density_distribution} to be $c_0 = c_1 = 1$ and $c_k = 0\:\: \forall k\geq 2$. By switching on the second coefficient $c_2$, we turn on the next-nearest neighbor hopping term $t_2$ in the Hamiltonian~\eqref{eq:H_eff_MB}. This corresponds to an effective triangular ladder geometry, see Fig.~\ref{fig1}(b), thus showing that our implementation allows for realizations beyond purely 1D systems. As mentioned in the previous Section, the relative strength and sign of both $t_1$ and $t_2$ is fully tunable by the ratio $c_2/c_1$ and $\varphi_{1,2}$, respectively. In such a simple way, we can then achieve regimes where geometrical frustration occurs by fixing $t_1,t_2>0$ or, as reported in Fig.~\ref{fig1}(b), $t_1<0$, $t_2>0$, see \cite{sato2011,Struck2013,greschner2013,mishra2015,Anisimovas2016,Cabedo2020,Barbiero2023,Roy2022,Halati2023,baldelli2024,burba2024}. It is important to remark that contrary to ultracold atomic systems where engineering frustration \cite{Struck2011,Struck2012} requires potentially detrimental fast periodic drivings, here frustration is created just through geometry manipulations. Moreover, large frustration in addition to the present weak interactions $U_{n,n'}$ can give rise to chiral superfluidity, see for instance \cite{greschner2013,Roy2022,Halati2023,baldelli2024,burba2024}, whose interest spans from high energy~\cite{Son2009} 
to condensed matter ~\cite{Sato_2017} systems. 
Interestingly, the range of hopping can be further extended by turning on $c_3$. In such a way, see Fig.~\ref{fig1}(c), we get an effective triangular frustrated system where up to tree effective lattice sites are connected. Crucially, this latter example represents the building block of purely two dimensional geometries displaying spin liquid phases \cite{Zhu2015,Balents2010}.
\paragraph{Power law tunneling processes---}
By switching on further coefficients $c_i$ and adjusting their value, one can engineer effective power-law decaying hopping terms
\begin{equation}
    t_{n,n'}\propto \frac{1}{|n-n'|^{\beta}}
\label{eq:H_power_law}
\end{equation}
with a power $\beta$ that can be freely tuned. In order to achieve that, one needs to tune the relative strength of the coefficients $c_i$ accordingly: $c_i \sim i^{-\beta}$. Figure~\ref{fig3} clearly shows that $\beta$ can acquire different values in a very controlled way so to reproduce both almost local, i.e $\beta\geq1$, and highly nonlocal, i.e $\beta\leq1$, regimes. In this regard, it is important to underline that while similar features have been achieved in trapped-ion quantum simulators mimicking  spin$-1/2$ models~\cite{PhysRevLett.103.120502,Britton_2012,Islam_2013,jurcevic2014,kokail2019}, in our case the tunability of $\beta$ applies to systems with an extended local Hilbert space.
\paragraph{Classical gauge fields---}

\begin{figure}[t]
\includegraphics[width=\columnwidth]{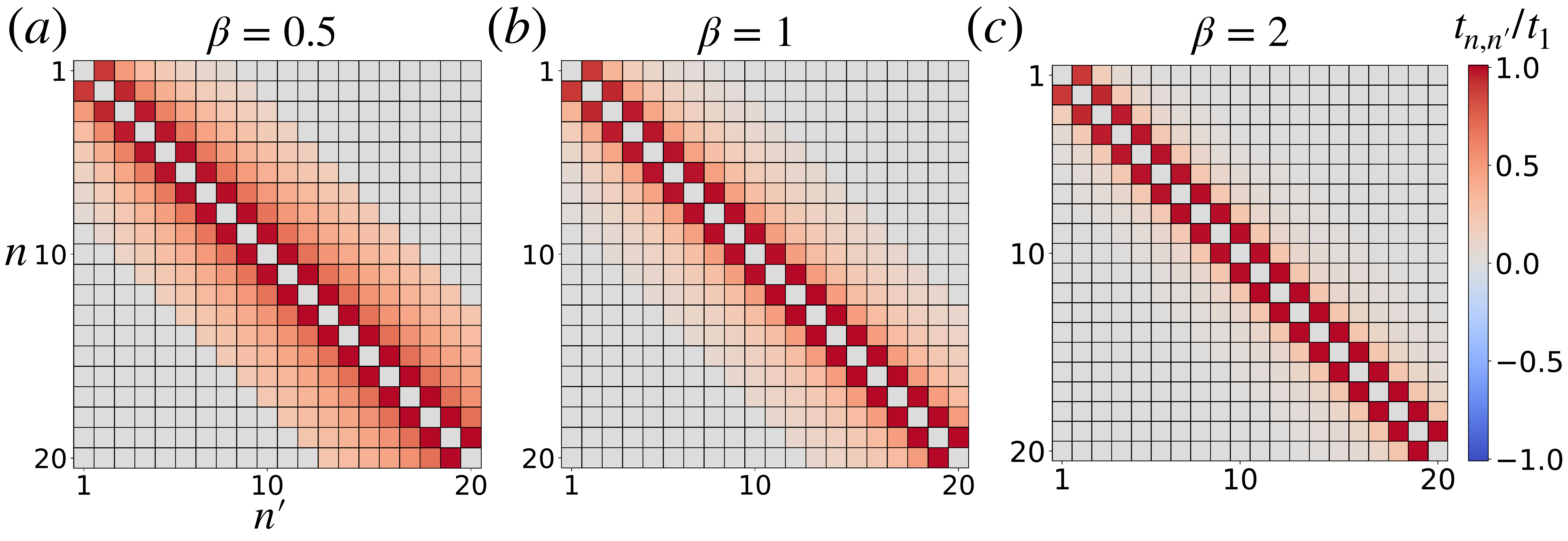}
\caption{\textbf{Power-law decay of the hopping amplitudes} The matrix $t_{i,j}$ representing the hopping amplitudes in the kinetic Hamiltonian as in Eq.~\eqref{eq:H_power_law} for three different values of the exponent: 
(a) $\beta = 0.5$,  (b) $\beta = 1$ and (c) $\beta = 2$. The first seven hopping amplitudes ($i-7\leq j \leq i+7$) are tuned to be nonzero by switching on the corresponding $c$ coefficients in the density of the scatterers.} 
%
\label{fig3}
\end{figure}
\begin{figure}[t]
\includegraphics[width=\columnwidth]{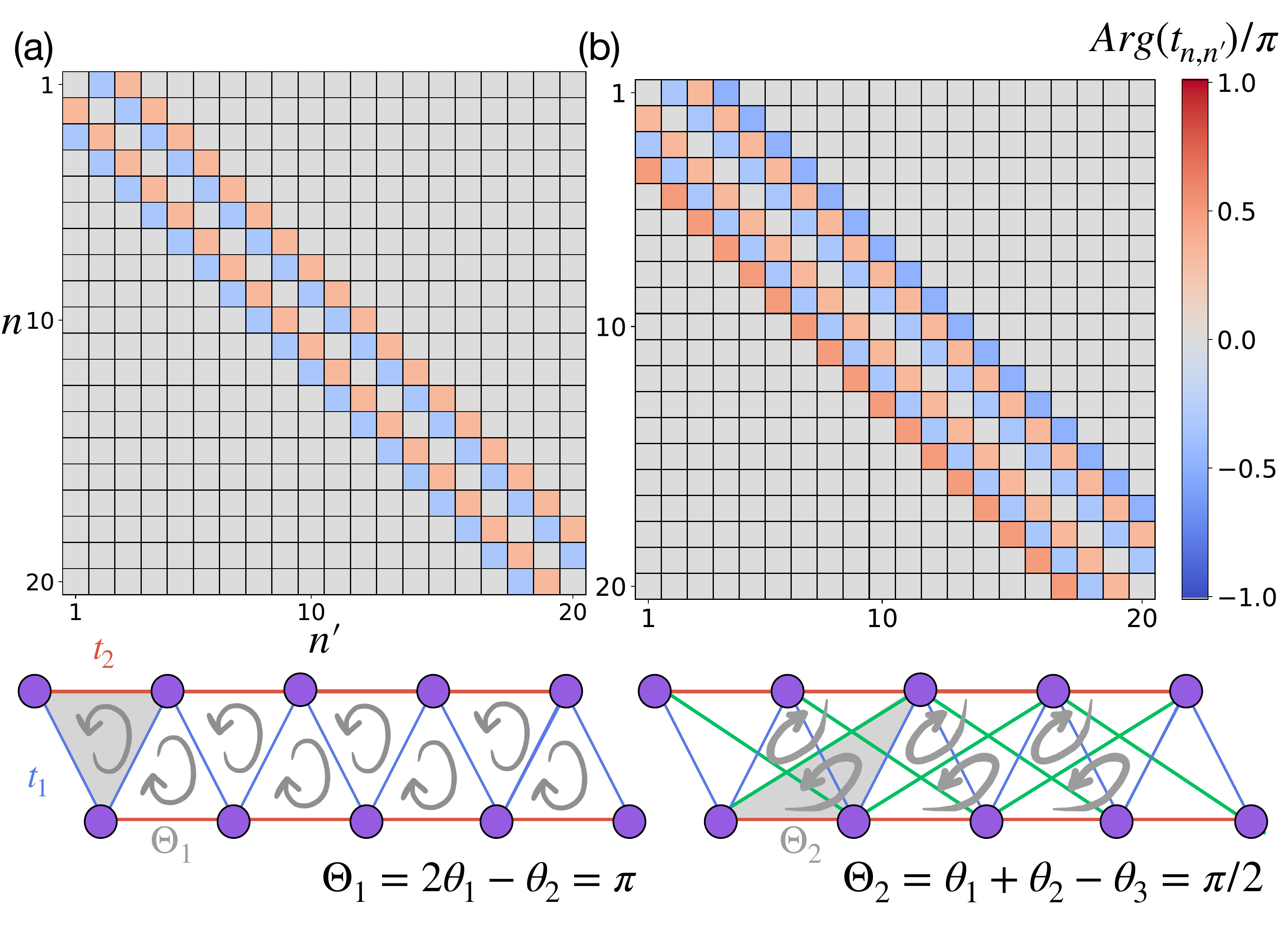}
\caption{\textbf{Classical gauge fields.}
(a) The complex phase of the nearest ($t_1$) and next-nearest ($t_2$) neighbor hopping can be tuned so that going around a plaquette on the triangular ladder, the bosonic particle acquires a phase of $\Theta_1 = \pi$.
(b) By switching on the next-to-next-nearest neighbor hopping ($t_3$), in addition to $\Theta_1$, another plaquette, defined by $t_1,t_2$ and $t_3$ can be identified. Choosing the complex phase of $t_3$ accordingly, a phase shift of $\Theta_2 = \pi/2$ by going around this second plaquette can be induced.}
\label{fig2}
\end{figure}
The versatility characteristic of our approach becomes even more spectacular when considering rotations of the hoppings in the complex plane. Here, as shown in Fig. \ref{fig2} artificial gauge fields take place. Contrary to other setups like cold atomic \cite{Dalibard2011,Aidelsburger2018}, solid state \cite{VOZMEDIANO2010109}, photonic \cite{HAFEZI2014} and optomechanical \cite{Walter_2016} settings,  
our setup allows for arbitrary adjusting of the Peierls phase by one control parameter--the relative orientation between the molecular cloud and the LG modes. More specifically, the Peierls phase acquired by the photon hopping from the $n$th LG mode to the $(n+k)$th LG mode is given by the sum of the azimuthal phase difference between the two modes and the phase $\varphi_k$. The ability to tune $\varphi_k$ arbitrarily by simply rotating the corresponding $\cos$ component in the density of the molecular cloud [see Eq.~\eqref{eq:density_distribution}] around the $z$ axis, allows us to engineer arbitrary translationally invariant Peierls phases for each hopping independently. 
In Fig.~\ref{fig2} (a) we show the complex phase of the kinetic part of the effective Hamiltonian, chosen so that in the triangular ladder geometry, a phase of $\Theta_1 = \pi$ is acquired by going around a plaquette. In Fig.~\ref{fig2} (b) we add one layer of complexity by introducing a next-to-next-nearest neighbor hopping with a complex amplitude, so that two different phases -- $\Theta_1 = \pi$ and $\Theta_2 = \pi/2$ are acquired, dependent on the triangle. Specifically here the generated gauge fields can become effectively long range as they can extended over different triangular plaquettes. As we underline, this unique feature can be made by controlling the density profile of the molecular cloud thus further proving the conceptual simplicity of our designed setup.
\paragraph{Conclusions and outlook---}
We studied the behavior of photons scattering off resonantly from a molecular cloud. In such a context, we derived a highly versatile Bose-Hubbard model where the elementary constituents are represented by the scattered photonic modes. Here, we demonstrated that the molecular distribution represents a powerful control parameter in order to achieve interesting configurations. Specifically, we first proved that, by properly shaping the molecular geometry, 
geometrical frustration naturally emerges. In this regard, we remark that configurations where geometrical frustration coexists with local and nonlocal interactions, as in our model, have been recently proposed as the ideal playground to explore states of matter characterized by chiral order \cite{burba2024} and exotic phase transitions \cite{baldelli2024}.
Inspired by recent results proving that power law tunneling processes with small decay exponent might represent a fundamental tool for quantum computing purposes \cite{Lewis2021,king2025}, we further demonstrated that long-range hopping amplitudes and their relative decay can be shaped at will.
Finally, we showed that for rotations of the molecular cloud 
both local and nonlocal gauge fields are acquired by scattered modes at each tunneling event. 
Crucially, this feature represents a fundamental tool
to create paradigmatic topological phases \cite{Haldane1988}.\\ 
We conclude by stressing that our work naturally motivates promising research directions. In particular, the search for specific molecules with a small gap and well-separated ground and first excited vibrational state from the rest of the spectrum can in principle open the path toward the exploration of strongly interacting photonic regimes \cite{chang2014,carusotto2020}. 
Our results thus represent an alternative and powerful approach to unveil novel quantum regimes induced by complex coupling structures. 
%
\paragraph{Acknowledgments---}
We warmly thank S. Asban for his contribution in the initial phase of this project. We also thank I. Carusotto and O. Firstenberg for discussions. ICFO-QOT group acknowledges support from:
European Research Council AdG NOQIA; MCIN/AEI (PGC2018-0910.13039/501100011033,  CEX2019-000910-S/10.13039/501100011033, Plan National FIDEUA PID2019-106901GB-I00, Plan National STAMEENA PID2022-139099NB, I00, project funded by MCIN/AEI/10.13039/501100011033 and by the “European Union NextGenerationEU/PRTR" (PRTR-C17.I1), FPI); QUANTERA MAQS PCI2019-111828-2;  QUANTERA DYNAMITE PCI2022-132919, QuantERA II Programme cofunded by European Union’s Horizon 2020 program under Grant Agreement No 101017733; Ministry for Digital Transformation and of Civil Service of the Spanish Government through the QUANTUM ENIA project call - Quantum Spain project, and by the European Union through the Recovery, Transformation and Resilience Plan - NextGenerationEU within the framework of the Digital Spain 2026 Agenda; Fundació Cellex; Fundació Mir-Puig; Generalitat de Catalunya (European Social Fund FEDER and CERCA program, AGAUR Grant No. 2021 SGR 01452, QuantumCAT \ U16-011424, cofunded by ERDF Operational Program of Catalonia 2014-2020); Barcelona Supercomputing Center MareNostrum (FI-2023-3-0024); Funded by the European Union (HORIZON-CL4-2022-QUANTUM-02-SGA  PASQuanS2.1, 101113690, EU Horizon 2020 FET-OPEN OPTOlogic, Grant No 899794, QU-ATTO, 101168628),  EU Horizon Europe Program (This project has received funding from the European Union’s Horizon Europe research and innovation program under grant agreement No 101080086 NeQSTGrant Agreement 101080086 — NeQST); ICFO Internal “QuantumGaudi” project; European Union’s Horizon 2020 program under the Marie Sklodowska-Curie grant agreement No 847648; P.P.P. acknowledges also support from the “Secretaria d’Universitats i Recerca del Departament de Recerca i Universitats de la Generalitat de Catalunya” under grant 2024 FI-3 00390, as well as the European Social Fund Plus. L. B. acknowledges financial support within the DiQut Grant No. 2022523NA7 funded by European Union– Next Generation EU, PRIN 2022 program (D.D. 104- 02/02/2022 Ministero dell’Università e della Ricerca).

Views and opinions expressed are however those of the author(s) only and do not necessarily reflect those of the European Union, European Commission, European Climate, Infrastructure and Environment Executive Agency (CINEA), or any other granting authority.  Neither the European Union nor any granting authority can be held responsible for them.

The data that support the findings of this article are openly available~\cite{popov_2025_14900650}.

\bibliography{ref}

\section{End Matter}

In this End Matter, we provide estimates on the order of magnitude of the relevant for the effective Hamiltonian processes with respect to the laser frequency. By analyzing how the magnitude of the hoppings and the interaction strength of the bosonic modes scale with the laser frequency and intensity and with the characteristics of the molecular transition, we show that there are regimes in which both hoppings and interactions are sizable. Furthermore, we estimate the order of magnitude of the deviation from the paraxial approximation and show that this holds in the regimes considered. 

\subsection{Order of magnitude for hopping term and interactions vs laser frequency}
For the light beam with intensity $I$ and frequency $\omega_0$, the order of the hopping amplitude can be estimated using Eq.~\eqref{eq:vector_potential} and Eq.~\eqref{eq:t_integral} and combining them as follows

\begin{align}
    t \propto \frac{e^2A^2_0}{2mc^2},
\end{align}
where with $A_0$ we denote the amplitude of the vector potential [for more details, see Eq.~(S13) in Supplemental Material]. This can be expressed through the intensity of the beam as follows
\begin{align}
    A^2_0 = E^2_0/\omega_0^2 = \frac{I}{2c\epsilon_0\omega_0^2}.
\end{align}
Substituting into the first equation, we obtain
\begin{align}
    t \propto \frac{e^2I}{4\epsilon_0mc^3\omega_0^2}.
\end{align}
We can estimate the order of magnitude by substituting the constants in SI units. Denoting $\omega_0 \sim 10^{p_{\omega_0}}$ Hz and $I \sim 10^{p_I} W/m^2$, we obtain for the ratio
\begin{align}
    \log R_t = \log\frac{t}{\hbar\omega_0} = 15+p_I-3p_{\omega_0}.
\end{align}
For realistic laser beams $p_I = 18$ and $p_{\omega_0} = 11$ (100 GHz)~\cite{Normand90}, therefore $\log R_t \sim 0$.
A similar calculation can be done for the interaction strength. Noting that the amplitude is $\propto t^2$, we write
\begin{align}
    R_U = \frac{\nu^2t^2}{\hbar\omega_0\Delta} = \nu^2\frac{t}{\hbar\omega_0}\frac{t}{\hbar\Omega}, 
\end{align}
where $\Delta = \hbar\Omega$. From the condition above, we see that $\log R_U\sim -2\log\nu - \log(t/\hbar\Omega) $. We do not have knowledge of typical values for the overlap $\nu$. However, the range of expected values would be $\nu\leq 0.1$. While the upper bound is surely optimistic, the ratio $R_U$ could still be sizable due to the fact that $\Omega < \omega_0$ and therefore the second $\log$ can be negative, counteracting the smallness of $\nu$.

\subsection{Paraxial approximation validity}
Here we estimate the order of magnitude of the corrections to the energy $\delta_n$ of each LG mode $n = (l,p)$. In the paraxial approximation, the first order correction to the energy of the beam mode is related to the transverse momentum $k_{\perp}$ as follows

\begin{align}
    \delta_n \propto \langle k^2_{\perp}\rangle/k_z,
\end{align}
where $k_z = \omega_0/c$ is the wave vector of the laser. Expressed in terms of the quantum numbers of the LG mode, the transversal momentum reads
\begin{align}
    \langle k^2_{\perp}\rangle = \frac{2(2p+|l|+1)}{w_0^2}
\end{align}
with $w_0$ being the beam waist. Putting both equations together, we get for the ratio of the energy shift and the laser frequency
\begin{align}
    \delta_n/\omega_0 \propto 2(2p+|l|+1)c^2/(w^2_0\omega_0^2).
\end{align}
Therefore, the energy shift is proportional to the ratio of the beam waist and the wavelength of the laser. Better paraxial approximations are reached for wider beams with shorter wavelength. As we mentioned above, in order for the relevant processes in the Hamiltonian to be sizable, we need a longer wavelength. Therefore, there is a range of values for which both the processes are sizable and the paraxial approximation is good. 

For the example above, $\omega_0 \sim 10^{11}$Hz and taking $w_0\sim 1$cm, we have $\delta_n/\omega \propto 10^{-2}$, which renders the paraxial approximation good on the timescale of the hopping.

\clearpage

\end{document}


\title{Supplemental Material for\\ ``Geometrical frustration, power law tunneling and non-local gauge fields from scattered light"}

\author{Pavel P. Popov}
\affiliation{ICFO - Institut de Ci\`encies Fot\`oniques, The Barcelona Institute of Science and Technology, 08860 Castelldefels (Barcelona), Spain}

\author{Joana Fraxanet}
\affiliation{ICFO - Institut de Ci\`encies Fot\`oniques, The Barcelona Institute of Science and Technology, 08860 Castelldefels (Barcelona), Spain}

\author{Luca Barbiero}
\affiliation{Institute for Condensed Matter Physics and Complex Systems,
DISAT, Politecnico di Torino, I-10129 Torino, Italy}

\author{Maciej Lewenstein}
\affiliation{ICFO - Institut de Ci\`encies Fot\`oniques, The Barcelona Institute of Science and Technology, 08860 Castelldefels (Barcelona), Spain}
\affiliation{ICREA, Pg. Lluis Companys 23, ES-08010 Barcelona, Spain}

\maketitle

In this Supplemental Material, we give the intermediate steps of the derivation of the effective Hamiltonian, that were skipped in the main text for purpose of saving space. We also give information about the spatial profile of Laguerre-Gaussian modes, the bosonic degrees of freedom in our model.

\section{Effective many-body Hamiltonian}
In this section, we systematically derive the effective Hamiltonian of Eq.~(8) in the main text. We write the Hamiltonian as

\begin{align}\label{eq:H1}
{\cal H} = \underbrace{\sum_\alpha \left(\frac{\vec{p}_\alpha}{2m} + V(\vec{r}_\alpha)\right) + \omega_0 \sum_n b_n^\dagger b_n}_{\:\:\:\:\:\:=: H_0} + \underbrace{\sum_\alpha \frac{e^2}{2mc^2} \vec{A}(\vec{r}_\alpha)^2 + \sum_n \delta_n b_n^\dagger b_n}_{\:\:\:\:\:\:=: H_1},
\end{align}
where $\delta_n = \omega_n - \omega_0$ are small corrections for the frequencies of the modes in the expansion of the EM part of the Hamiltonian. In our case, these modes are chosen to be eigenmodes of the paraxial Helmholtz equation~\cite{Aiello2005}. 

Assuming that each molecule is a two-level system with ground state $\ket{g_{\alpha}}$ and excited state $\ket{e_{\alpha}}$, and a gap $E = \Delta$, 
the unperturbed part $H_0$ can easily be diagonalized and the Hamiltonian takes the form 

\begin{align}\label{eq:perturbation}
{\cal H}_0 &= \sum_\alpha \Delta \vert e_\alpha \rangle \langle e_\alpha \vert + \omega_0 \sum_n b_n^\dagger b_n, \\
{\cal H}_1 &= \sum_n \delta_n b_n^\dagger b_n + \sum_\alpha \frac{e^2}{2mc^2}   \vec{A}(\vec{r}_\alpha)^2.  
\end{align}
The EM part of the Hamiltonian is diagonalized by eigenmodes of the Maxwell equations, and in the following, we work in the sector of constant photon number. Our theoretical considerations go beyond the dipolar approximation, and therefore, the vector potential acts non-trivially on the molecular eigenstates. We proceed by calculating the effective Hamiltonian for the EM modes in second order perturbation theory, where the perturbation Hamiltonian is ${\cal H}_1$.

\subsection{Hopping term amplitudes}

The first order correction to the energy is given by 
\begin{align}
&\frac{e^2}{2mc^2} \sum_{\alpha}\bra{g_{\alpha}}\vec{A}(\vec{r}_\alpha)^2\ket{g_{\alpha}} + \sum_n\delta_nb^{\dagger}_nb_n = \frac{e^2}{2mc^2} \sum_{\alpha,n,m}\big(\bra{g_{\alpha}}f_{n}(\vec{r}_{\alpha})f^*_{m}(\vec{r}_{\alpha})\ket{g_{\alpha}}b^{\dagger}_nb_{m} + \text{h.c.}\big)+ \sum_n\delta_nb^{\dagger}_nb_n\notag\\ &\equiv \sum_{n,m}\big(t_{nm}b^{\dagger}_nb_{m}+\text{h.c.}\big)+ \sum_n\delta_nb^{\dagger}_nb_n,
\end{align}
where we defined 
\begin{align}
    t_{nm} = \frac{e^2}{2mc^2} \sum_{\alpha}\bra{g_{\alpha}}f_{n}(\vec{r}_{\alpha})f^*_{m}(\vec{r}_{\alpha})\ket{g_{\alpha}}
\end{align}
We further switch from discrete description of the molecules to a continuous one, $\sum_{\alpha} \rightarrow \int_Vd^3r\rho(\vec{r})$. The volume $V$ is determined by the total volume of the molecular cloud. The coefficients $t_{nm}$ become
\begin{align}
    t_{nm} = \frac{e^2}{2mc^2}\int_Vd^3r\rho(\vec{r})f_{n}(\vec{r})f^*_{m}(\vec{r}).
\label{eq:t_nm_1st_order}
\end{align}
Therefore, the diagonal part of the perturbation to the energy is 
\begin{align}
    {\cal H} = \sum_nt_{nn}b^{\dagger}_nb_n + \sum_n\delta_nb^{\dagger}_nb_n \equiv \sum_n\mu_nb^{\dagger}_nb_n
\label{eq:diag_1st_order}
\end{align}
and the off-diagonal, which corresponds to the hopping amplitudes between different modes, 
\begin{align}
    {\cal H} = \sum_{n,m}t_{nm}b^{\dagger}_nb_m + \text{h.c.}.
\label{eq:off_diag_1st_order}
\end{align}

\subsection{Density-density interactions}

The second order correction to the energy is given by
\begin{align}
-\frac{1}{\Delta}\Bigg(\frac{e^2}{2mc^2}\Bigg)^2\sum_{\alpha}|\bra{g_{\alpha}}\vec{A}(\vec{r}_\alpha)^2\ket{e_{\alpha}}|^2=  -\frac{1}{\Delta}\Bigg(\frac{e^2}{2mc^2}\Bigg)^2\sum_{\alpha}|\bra{g_{\alpha}}\sum_{n,m}f_{n}(\vec{r}_{\alpha})f^*_{m}(\vec{r}_{\alpha})\ket{e_{\alpha}}|^2
\end{align}
As we mentioned in the main text, we now rewrite the off-diagonal element of the vector potential as proportional to the diagonal one with proportionality factor $\nu$; using Eq.~(6) in the main text we obtain
\begin{align}
\sum_{\alpha}|\bra{g_{\alpha}}\vec{A}(\vec{r}_\alpha)^2\ket{e_{\alpha}}|^2 &=\nu^2\sum_{\alpha}|\bra{g_{\alpha}}\sum_{n,m}f_{n}(\vec{r}_{\alpha})f^*_{m}(\vec{r}_{\alpha})b^{\dagger}_nb_m + \text{h.c.}\ket{g_{\alpha}}|^2\notag\\
& = \nu^2\sum_{\alpha}\bra{g_{\alpha}}\sum_{n,m,n',m'}f_{n}(\vec{r}_{\alpha})f^*_{m}(\vec{r}_{\alpha})f_{n'}(\vec{r}_{\alpha})f^*_{m'}(\vec{r}_{\alpha}) b^{\dagger}_{n}b_{m}b^{\dagger}_{n'}b_{m'}\ket{g_{\alpha}}\notag\\&+\nu^2\sum_{\alpha}\bra{g_{\alpha}}\sum_{n,m,n',m'}f_{n}(\vec{r}_{\alpha})f^*_{m}(\vec{r}_{\alpha})f^{*}_{n'}(\vec{r}_{\alpha})f_{m'}(\vec{r}_{\alpha})b^{\dagger}_{n}b_{m}b_{n'}b^{\dagger}_{m'}\ket{g_{\alpha}}\notag\\& + \text{h.c.}
\label{eq:interaction_calculation}
\end{align}
Replacing the discrete description of the molecular cloud with a continuous one, we obtain for the second order correction
\begin{align}
-\frac{1}{\Delta}\Bigg(\frac{e^2}{2mc^2}\Bigg)^2\sum_{\alpha}|\bra{g_{\alpha}}\vec{A}(\vec{r}_\alpha)^2\ket{e_{\alpha}}|^2 =-\sum_{m,n}U_{mn}(4b^{\dagger}_n b_nb^{\dagger}_m b_m+3b^{\dagger}_n b_n), 
\label{eq:interaction_result}
\end{align}
where we defined the interaction strength 

\begin{align}
    U_{mn} = \frac{\nu^2}{\Delta}\Bigg(\frac{e^2}{2mc^2}\Bigg)^2\int_Vd^3\vec{r}\rho(\vec{r})|f_n(\vec{r})|^2|f_m(\vec{r})|^2.
\end{align}
It should be noted that the right-hand side of Eq.~\eqref{eq:interaction_calculation} contains additional terms beyond the ones in Eq.~\eqref{eq:interaction_result}. However, these additional terms can be neglected since the LG modes carry a non-trivial time-dependent phase factor due to their small energy shifts $\delta_{n}$. Therefore, only those terms survive a time averaging, for which these phase factors exactly cancel out.

\section{Laguerre-Gaussian modes}
As we explained in the main text, the set of EM eigenmodes used in this Letter is the Laguerre-Gaussian modes, due to the cylinder symmetry of the considered setup. The spatial profile of these modes is given by 

\begin{align}
    f_{l,p}(\vec{r}) = &\bigg(\frac{\hbar}{8\pi^2\epsilon_0\omega_0(1+\vartheta^4)^{1/2}L}\bigg)^{1/2}\frac{C_{l,p}}{w(z)}\bigg(\frac{r\sqrt{2}}{w(z)}\bigg)^{|l|}\exp{\bigg(-\frac{r^2}{w^2(z)}\bigg)}L^{|l|}_{p}\bigg(\frac{2r^2}{w^2(z)}\bigg)\exp{\bigg(-ik_0\frac{r^2}{2R(z)}\bigg)}\exp{(-il\phi)}\exp{[i\psi(z)]},
\label{eq:LGmodes_formula}
\end{align}
where $w(z)$ is the waist radius, $R(z)$ the radius of curvature, $k = 2\pi/\lambda$ the wave-number for the wavelength $\lambda$ and $\psi(z)$ is the Gouy phase. In the prefactor, the $\epsilon_0$ is the vacuum dielectric permitivity, $\vartheta$ is the (small) divergence angle of the Gaussian beam and $L$ is the extent of the cavity along the $z-$direction. The coefficient $C_{l,p}$ is a normalization constant 
\begin{align}
    C_{l,p} = \sqrt{\frac{2p!}{\pi(p + |l|)!}},
\end{align}
so that the orthogonality condition
\begin{align}
    \int d^3\vec{r}f_{l,p}(\vec{r})f_{l',p'}(\vec{r}) = \delta_{l,l'}\delta_{p,p'}
\end{align}
and the normalization condition 
\begin{align}
    \int d^3\vec{r}|f_{l,p}(\vec{r})|^2 = 1
\end{align}
hold. We simplify the description by assuming small divergence of the beam along the $z$-direction and therefore, in cylindrical coordinates, the integrals defining the coefficients of the effective Hamiltonian along that direction are rather trivial (see Eqns.~(9)-(11) in the main text). The integration in the plane perpendicular to $z$ is non-trivial and we perform it numerically.

\bibliography{ref}